\PassOptionsToPackage{table}{xcolor}

\documentclass[unnumsec,webpdf,contemporary,large]{oup-authoring-template}%

\usepackage{hyperref}
\usepackage{natbib}
\usepackage{makecell}

\setcitestyle{round}

\usepackage[table]{xcolor}

\graphicspath{{Fig/}}

\usepackage[utf8]{inputenc}
\usepackage[margin=1in]{geometry}
\usepackage{booktabs}
\usepackage{multirow}
\theoremstyle{thmstyleone}%
\theoremstyle{thmstyletwo}%
\theoremstyle{thmstylethree}%

\begin{document}

\journaltitle{Journal Title Here}
\DOI{DOI HERE}
\copyrightyear{2022}
\pubyear{2019}
\access{Advance Access Publication Date: Day Month Year}
\appnotes{Paper}

\firstpage{1}


\title[FASTR Sequencing Data Format]{FASTR: Reimagining FASTQ via Compact Image-inspired Representation}

\author[1,2]{Adrian Tkachenko}
\author[1,2]{Sepehr Salem}
\author[2]{Ayotomiwa Ezekiel Adeniyi}
\author[3]{Zülal Bingöl}
\author[1,2]{Mohammed Nayeem Uddin}
\author[1]{Akshat Prasanna}
\author[2,5]{Alexander Zelikovsky}
\author[4,5]{Serghei Mangul}
\author[3]{Can Alkan}
\author[1,2,$\ast$]{Mohammed Alser\ORCID{0000-0002-6117-3701}}

\authormark{Tkachenko et al.}

\address[1]{\orgdiv{ALSER Lab, Computational Life Sciences}, \orgname{Georgia State University}, \orgaddress{Atlanta, GA 30303, USA}}
\address[2]{\orgdiv{Department of Computer Science}, \orgname{Georgia State University}, \orgaddress{Atlanta, GA 30303, USA}}
\address[3]{\orgdiv{Department of Computer Engineering}, \orgname{Bilkent University}, \orgaddress{Ankara 06800, Turkey}}
\address[4]{\orgname{Sage Bionetworks}, \orgaddress{Seattle, WA, USA}}

\address[5]{\orgdiv{Department of Biological and Morphofunctional Sciences, College of Medicine and Biological Sciences}, \orgname{Stefan cel Mare University of Suceava}, \orgaddress{720229 Suceava, Romania}}

\corresp[$\ast$]{Corresponding author. \href{mailto:malser@gsu.edu}{malser@gsu.edu}}
\received{Date}{0}{Year}
\revised{Date}{0}{Year}
\accepted{Date}{0}{Year}

\abstract{
\textbf{Motivation}: High-throughput sequencing (HTS) enables population-scale genomics but generates massive datasets, creating bottlenecks in storage, transfer, and analysis. FASTQ, the standard format for over two decades, stores one byte per base and one byte per quality score, leading to inefficient I/O, high storage costs, and redundancy. Existing compression tools can mitigate some issues, but often introduce costly decompression or complex dependency issues. \\
\textbf{Results}: We introduce \emph{\textbf{FASTR}}, a lossless, computation-native successor to FASTQ that encodes each nucleotide together with its base quality score into a single 8-bit value. FASTR reduces file size by at least 2$\times$ while remaining fully reversible and directly usable for downstream analyses. Applying general-purpose compression tools on FASTR consistently yields higher compression ratios, 2.47, 3.64, and 4.8$\times$ faster compression, and 2.34, 1.96, 1.75$\times$ faster decompression than on FASTQ across Illumina, HiFi, and ONT reads. 
FASTR is machine-learning-ready, allowing reads to be consumed directly as numerical vectors or image-like representations. We provide a highly parallel software ecosystem for FASTQ–FASTR conversion and show that FASTR integrates with existing tools, such as minimap2, with minimal interface changes and no performance overhead. By eliminating decompression costs and reducing data movement, FASTR lays the foundation for scalable genomics analyses and real-time sequencing workflows.\\
\textbf{Availability and Implementation:} \href{https://github.com/ALSER-Lab/FASTR}{https://github.com/ALSER-Lab/FASTR}
}

\keywords{FASTQ, FASTR, genome sequence analysis}


\maketitle

\section{Introduction}



Advancing genomic analyses is now a critical imperative. The challenge is to move beyond mere accuracy and achieve rapid, efficient genomic interpretation for entire populations \citep{alser2022molecules}. Modern high-throughput sequencing (HTS) technologies are instrumental in addressing this challenge. These systems generate enormous volumes of data (called reads) stored in FASTQ format, which broadly includes three types of information: genomic sequence, base quality scores, and metadata (called a header) associated with each read. Genomic sequences, known as reads, are typically produced in quantities ranging from millions to billions, are randomly sampled from the genome, and span lengths from hundreds to millions of base pairs. Each read is accompanied by a base quality sequence that reflects the HTS system’s confidence in identifying individual bases. The throughput of prominent systems from Illumina, PacBio, and Oxford Nanopore is constantly improving. The scale of existing sequencing data is already staggering, with public repositories containing tens of petabases (NIH’s SRA stores over $32 \times 10^{15}$ bytes). As sequencing technologies advance and large-scale initiatives expand, this data reservoir will continue its trajectory of exponential growth, outpacing advances in computational power~\citep{katz2022sequence, alser2025taming}.

Therefore, it is crucial to store this vast amount of sequencing data in efficient, compact file formats that support fast reading, parsing, and writing. The most widely used FASTQ format~\citep{cock2010sanger}, introduced in the late 1990s at the Sanger Institute as an extension of the FASTA format, was designed to store one byte per base and one byte per corresponding sequencing quality score. At the time of its introduction, read lengths and throughput were modest (e.g., approximately 600 bases and 200 reads per day on the ABI 377), making a human-readable text format a reasonable choice. Today, however, FASTQ files can easily exceed 50 GB in size, with read lengths ranging from hundreds to millions of bases, rendering the traditional format increasingly inefficient. Additionally, FASTQ stores redundant information while omitting other important details, resulting in wasted space and potential ambiguity. 
Each read record contains two header lines and often repeats the same information (e.g., accession number and instrument name) across all reads in the file. At the same time, the format lacks explicit information about what is being sequenced, the sequencing protocol or run, and the standard used for the quality score encoding.

There are many alternative approaches that aim to replace or improve the FASTQ format. These approaches can be divided into four main categories: 1) Introducing alternative formats, such as CRAM~\citep{fritz2011efficient}, SAM~\citep{li2009sequence} (its compressed version, BAM), and SRA~\citep{leinonen2011sequence}. 2) Applying general-purpose compression tools, such as gzip, to reduce FASTQ storage footprint. 3) Developing specialized compression tools, such as ReNANO~\citep{dufort2021renano}, SCALCE~\citep{hach2012scalce}, FQSqueezer~\citep{deorowicz2020fqsqueezer}, and SPRING~\citep{chandak2019spring}, that compress each line in FASTQ differently. 4) Reducing FASTQ data through subsampling, deduplication, sorting, trimming, and quality-score quantization. Each category has limitations. Alternative formats often depend on a reference genome and/or incur significant computational overhead. General-purpose compression requires decompression (to disk or via streaming) back to FASTQ prior to data parsing, which increases parsing time by 2--10$\times$~\citep{Li2020FastLanguages}. Specialized compression tools are frequently sequencing-platform dependent and likewise require computationally-expensive decompression to FASTQ. Data-reduction approaches are lossy and rely on heuristics whose effects on downstream analyses can be difficult to predict.

Our goal is to eliminate all existing limitations of the FASTQ format and enable efficient storage and processing of sequencing data. To this end, we introduce \textit{\textbf{FASTR}}, a highly optimized and lossless file format for sequencing data that naturally supports fast reading, parsing, and writing without the need for compression/decompression. FASTR is based on two key ideas: 1) correlating the base quality information to their corresponding bases and encoding both information as a 1-byte pixel in a grayscale image, 2) eliminating redundancy in each read header and storing them only once in a global file header.
A pixel in a grayscale image can represent integer values in the range 0-255, which is wide enough to encode 5 bases of DNA and RNA (i.e., A, C, G, T/U, and N) along with 63 base quality scores. This can be achieved in different ways, which we will discuss in detail in the \hyperref[sec:methods]{Methods} section. As sequencing technology accuracy improves, the range of base quality values reduces, leaving more room to encode additional metadata.

The choice of an image-inspired format is naturally supported and generated by modern HTS technologies (e.g., Illumina, PacBio, BGI/MGI, and GenapSys), where bases and their quality scores are already light intensities captured in images.
This provides several key benefits such as: 1) facilitating format conversions from raw sequencing data into FASTR format, 2) enabling real-time processing of sequencing data as they are generated, 3) enabling early filtration of low-quality data before completing the sequencing run, and 4) enabling direct compatibility with machine learning pipelines through its numeric representation.

The contributions of this paper are as follows:
\begin{enumerate}
    \item We introduce \textit{\textbf{FASTR}}, a lossless and computation-native successor to the FASTQ format that represents sequencing reads as compact numerical integers, enabling fast reading, parsing, and writing without requiring decompression.
    \item We propose a novel implicit partitioned range encoding scheme that encodes each nucleotide together with its base quality score into a single 8-bit grayscale value, achieving more than a 2$\times$ reduction in file size compared to FASTQ while preserving all information. Applying general-purpose compression tools (e.g., gzip) on FASTR consistently yields higher compression ratios and 2.47$\times$, 3.64$\times$, and 4.8$\times$ faster compression than that on FASTQ for Illumina, HiFi, and ONT reads, respectively.
    \item We design a flexible and extensible FASTR file header specification that captures essential sequencing experiment metadata across major platforms, including Illumina, PacBio, and Oxford Nanopore, as well as commonly used Phred quality score standards.
    \item We develop a highly parallel software ecosystem for FASTQ–FASTR conversion and demonstrate that FASTR can be efficiently consumed by existing downstream tools, such as minimap2~\citep{li2018minimap2}, with minimal interface-level code changes and no measurable performance overhead.
\end{enumerate}

\section{Methods}
\label{sec:methods}

\subsection{Overview}
\label{subsec:overview}

\begin{figure*}[!t]
    \centering
    \includegraphics[width=0.8\textwidth]{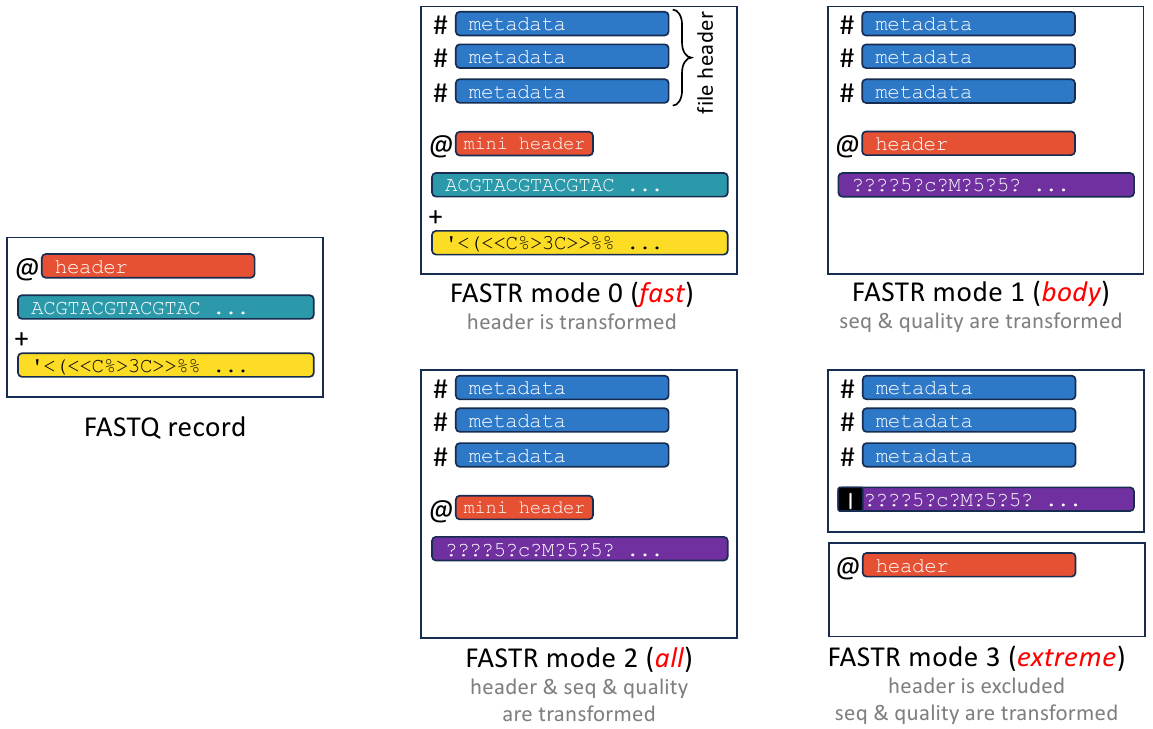}
    \caption{Overview of FASTR processing modes compared to a FASTQ record}
    \label{fig:overview-fastr}
\end{figure*}

The FASTR file format is designed to accelerate sequencing data analysis by minimizing I/O costs, storage footprint, and parsing time. It is a compact file format that represents sequencing reads as numerical vectors, allowing efficient parsing without the need for computationally expensive decompression. 
The FASTR format is composed of two elements: (1) a global file header containing metadata and decoding parameters, and (2) a sequence of read records storing numerical vectors that encode both bases and quality scores using the uint8 data type. The global file header reduces redundant metadata and preserves essential sample information that is typically omitted in FASTQ. FASTR supports four FASTQ-to-FASTR conversion modes: \textit{mode 0, mode 1, mode 2,} and \textit{mode 3}, as illustrated in Figure \ref{fig:overview-fastr}. Each mode is tailored to meet distinct requirements for storage efficiency, parsing performance, and information retention. In all modes, the global file header is retained. \textit{Mode 0} stores each read record in four lines identical to FASTQ, except that the read header is optimized to remove shared content with other read headers. \textit{Mode 1} represents each read with two lines: the original header and a numerical vector. \textit{Mode 2} follows the structure of \textit{mode 1} but uses the optimized header from \textit{mode 0}. \textit{Mode 3} stores only the numerical vector for each read, while headers are written to a separate file to enable conversion back to FASTQ when needed.

The FASTR pipeline consists of three main stages. First, the algorithm requires a standardized and validated input FASTQ file to ensure the correctness of numerical encoding. It analyzes and processes the FASTQ read headers to identify and reduce redundancy, optionally applying dictionary-based compression and removing repeated metadata. Second, the genomic bases and quality scores are converted to numerical values using our new partitioned encoding scheme, yielding a compact numeric vector representation. Because each base and its quality score are encoded into a single numeric value, FASTQ-to-FASTR conversion can be efficiently parallelized using chunk-based multiprocessing on modern multi-core CPUs. This enables FASTR to scale to large datasets while maintaining high throughput and low memory overhead. These independent numeric values can also be quickly parsed (e.g., using a lookup table) during downstream analysis, enabling existing FASTQ-compatible tools to adapt to FASTR with minimal changes (as we experimentally demonstrate in \hyperref[sec:results]{Results} section.

\subsection{File Header and Metadata Encoding}
\label{subsec:fastr-file-header}

We introduce a standardized file header for the FASTR file format to address two important limitations in FASTQ by systematically 1) explicitly recording commonly missing metadata and 2) eliminating all repeated substrings from individual read records. Conventional FASTQ files either omit critical provenance information (e.g., sample identifier, run date, flow cell, quality alphabet) or redundantly encode such metadata in every read header, inflating file size and complicating downstream parsing. For example, FASTQ headers frequently repeat identical string literals across millions of reads, differing only in small numeric fields such as coordinates or read indices, as highlighted in Table \ref{table:header-structure}. 

FASTR file header lines in all modes precede the first read record and follow a simple {\tt \#KEY=VALUE} syntax. This allows FASTR parsers to unambiguously distinguish file headers from read records, which begin with {\tt @} in both standard FASTQ and FASTR. The header is placed at the beginning of the file so that decoding parameters and metadata are available before any read records are processed. The file header captures three important types of metadata across at least 18 lines: first, it provides metadata describing the sequencing run, as shown in lines 3-9 of Table \ref{table:example-header}. Second, it provides metadata describing the properties of the sequencing data, such as lines 2, 10, and 15 in Table \ref{table:example-header}. Third, it provides metadata needed for reliable FASTR-to-FASTQ conversion, as shown in other lines of Table \ref{table:example-header}. To extract these three metadata types, FASTR separates each read header into two components: a static template and variable identifiers. The static template captures the invariant metadata common to all reads in a sequencing run, such as instrument identifiers and run descriptors. The variable identifiers capture the per-read fields, such as coordinates or indices. The static template is stored once in the FASTR file header, while only the variable identifiers are stored for each read. The {\tt STRUCTURE} field in the file header stores such a template for read identifiers using fixed literals and named placeholders (e.g., {\tt\{REPEATING\_1\}}, {\tt\{REPEATING\_2\}}), allowing only the truly variable components to appear in each read entry. FASTR also stores some parameters used for the FASTQ-to-FASTR conversion in the file header, enabling deterministic, lossless reconstruction of the original FASTQ file. Parsers can ignore unknown keys, validate critical fields (e.g., date formats, quality score alphabets), and extend the file header with additional lines when needed.

\begin{table*}[]
\centering
\small
    \begin{minipage}{\textwidth}
        \caption{FASTQ read header structure for different standards}
        \label{table:header-structure}
        
        \begin{tabular}{|l|l|} 
        \hline 
        \textbf{Standard} & \textbf{Header format structure (black: static template, \textcolor{red}{red: variable identifier}, \textcolor{blue}{blue: paired-end})} \\ 
        \hline 
        \textbf{Illumina} \tiny{(Casava 1.8+)} & @\textless{}instrument\textgreater{}:\textless{}run\textgreater{}:\textless{}flowcell\textgreater{}:\textcolor{red}{\textless{}lane\textgreater{}:\textless{}tile\textgreater{}:\textless{}x\textgreater{}:\textless{}y\textgreater} \textcolor{blue}{\textless{}read\textgreater{}}:\textless{}is\_filtered\textgreater{}:\textless{}control\textgreater{}:\textless{}index\textgreater{} \\
        \textbf{Illumina\_SRA}          & @\textless{}sra\_accession\textgreater{}.\textcolor{red}{\textless{}read\_number\textgreater} \textless{}original\_illu\textcolor{red}{mina\_header\textgreater} length=\textless{}length\textgreater{} \\
        \textbf{Illumina\_Old}          & @\textless{}instrument\textgreater{}:\textcolor{red}{\textless{}lane\textgreater{}:\textless{}tile\textgreater{}:\textless{}x\textgreater{}:\textless{}y\textgreater}{}\#\textless{}index\textgreater{}\textcolor{blue}{/\textless{}read\textgreater}{} \\
        \textbf{PacBio HiFi}            & @\textless{}instrument\_dateYYYYMMDD\_time\textgreater{}/\textcolor{red}{\textless{}zmw\textgreater}{}/ccs \textcolor{red}{extra tags} \\
        \textbf{PacBio\_HiFi\_SRA}      & @\textless{}sra\_accession\textgreater{}.\textcolor{red}{\textless{}read\_number\textgreater} \textless{}original\_pac\textcolor{red}{bio\_header\textgreater} length=\textless{}length\textgreater{} \\
        \textbf{PacBio CLR}             & @\textless{}instrument\_dateYYYYMMDD\_time\textgreater{}/\textcolor{red}{\textless{}zmw\textgreater{}/\textless{}start\textgreater{}\_\textless{}end\textgreater}{} \\
        \textbf{PacBio\_CLR\_SRA}       & @\textless{}sra\_accession\textgreater{}.\textcolor{red}{\textless{}read\_number\textgreater} \textless{}original\_pac\textcolor{red}{bio\_header\textgreater} length=\textless{}length\textgreater{} \\
        \textbf{ONT}                    & @\textcolor{red}{\textless{}uuid\textgreater runid=\textless{}runid\textgreater read=\textless{}read\textgreater ch=\textless{}channel\textgreater start\_time=\textless{}timestamp\textgreater}{} \\
        \textbf{ONT\_SRA}               & @\textless{}sra\_accession\textgreater{}.\textcolor{red}{\textless{}read\_number\textgreater \textless{}original\_ont\_header\textgreater} length=\textless{}length\textgreater{} \\
        \hline 
        \end{tabular}
    \end{minipage}                                                
\end{table*}

\begin{table*}[]

\centering
\small
    \begin{minipage}{\textwidth}

        \caption{An example of FASTR’s file header}
        \label{table:example-header}
    
        \begin{tabular}{|c|l|l|}
        \hline
        {\color[HTML]{000000} \textbf{\#}} & \textbf{Header example}                                                          & \textbf{Description of each line}                                          \\ \hline
        {\color[HTML]{F340CA} \textbf{1}}  & {\tt \#MODE=0}                                                  & {\tt FASTR mode: 0, 1, 2, or 3}                           \\
        {\color[HTML]{F340CA} \textbf{2}}  & {\tt \#SEQ-TYPE=pacbio\_hifi\_sra}                              & {\tt Data type, e.g., Illumina, PacBio, ONT, or SRA} \\
        {\color[HTML]{F340CA} \textbf{3}}  & {\tt \#ACCESSION=ERR15998594}                                   & {\tt NCBI accession number}                               \\
        {\color[HTML]{F340CA} \textbf{4}}  & {\tt \#INSTRUMENT=m84158}                                       & {\tt Sequencing instrument code}                          \\
        {\color[HTML]{F340CA} \textbf{5}}  & {\tt \#SAMPLE-ID=}                                              & {\tt Biological sample identifier}                        \\
        {\color[HTML]{F340CA} \textbf{6}}  & {\tt \#RUN-NUMBER=}                                             & {\tt Sequencing run identifier}                           \\
        {\color[HTML]{F340CA} \textbf{7}}  & {\tt \#RUN-DATE(YYYYMMDD)=251015}                               & {\tt Sequencing run date}                                 \\
        {\color[HTML]{F340CA} \textbf{8}}  & {\tt \#RUN-START-TIME=170119}                                   & {\tt Sequencing run start time}                           \\
        {\color[HTML]{F340CA} \textbf{9}}  & {\tt \#FLOW-CELL=}                                              & {\tt Flow cell identifier}                                \\
        {\color[HTML]{F340CA} \textbf{10}} & {\tt \#PHRED-ALPHABET=PHRED\_94}                                & {\tt Phred quality type}                                  \\
        {\color[HTML]{F340CA} \textbf{11}} & {\tt \#GRAY\_VALS={[}0{]},{[}3{]},{[}66{]},{[}129{]},{[}192{]}} & {\tt Partitioned range encoding information}              \\
        {\color[HTML]{F340CA} \textbf{12}} & {\tt \#LENGTH=y}                                                & {\tt Is sequence length included in the read header?}     \\
        {\color[HTML]{F340CA} \textbf{13}} & {\tt \#SECOND\_HEAD=0}                                          & {\tt Is the header repeated again after the ‘+’ sign?}    \\
        {\color[HTML]{F340CA} \textbf{14}} & {\tt \#SAFE\_MODE=}                                             & {\tt Is FASTR’s safe mode activated?}                     \\
        {\color[HTML]{F340CA} \textbf{15}} & {\tt \#PAIRED-END=}                                             & {\tt Is the data paired-end?}                                                    \\
        {\color[HTML]{F340CA} \textbf{16}} & {\tt \#PAIRED-END-SAME-FILE=}                                   & {\tt Is paired-end data in the same file?}                \\
        {\color[HTML]{F340CA} \textbf{17}} & {\tt \#QUAL\_SCALE=1 + 62*(ln(x+1)/ln(94))}                     & {\tt Equation used for base quality scaling}     \\
        {\color[HTML]{F340CA} \textbf{18}} & {\tt \#STRUCTURE=}                    & {\tt Template used to reconstruct the read header}  \\ 
        {\color[HTML]{F340CA} } & {\tt ERR15998594.\{REPEATING\_1\}m84158\_251015\_170119\_s2/\{REPEATING\_2\}/ccs} & {\tt} \\
        \hline
    \end{tabular}

    \end{minipage}                                                                                                                      
        
\end{table*}


\subsection{Individual Read Headers}
\label{subsec:ind-read-headers}

After the file header, FASTR stores all read records. A read record in FASTR can be formed using 1, 2, or 4 lines, depending on the chosen processing mode (mode 3, modes 1 or 2, and mode 0, respectively). In modes 0 and 2, FASTR stores an optimized version of the read header that contains only the variable identifiers determined during file header calculation (as illustrated in Table \ref{table:header-structure}). In mode 1, FASTR stores the same read headers as the original FASTQ file. This can be helpful for conservative users who prefer to keep the header unoptimized. In mode 3, FASTR excludes all read headers and stores them in a separate plain-text file. This could be exploited by parsers and downstream analysis tools to shorten their I/O overhead by reading less input data for applications that do not track read identifiers in their output, such as taxonomy profiling~\citep{meyer2022critical, liu2025analysis}, composition analysis (e.g., Sourmash~\citep{brown2016sourmash}), and quantifying abundances of transcripts (e.g., Kallisto~\citep{bray2016near}).

\subsection{Implicit Partitioned Range Encoding}
\label{subsec:read-rec}

After building read headers, FASTR encodes each genomic base together with its corresponding quality score into a single 8-bit numeric value. This reduces storage from two bytes per base–quality pair in standard FASTQ files to one byte, yielding an effective 2x reduction in size. The key idea behind FASTR encoding is to implicitly encode the base identity while explicitly encoding the quality score. Rather than dedicating fixed bit fields for bases and base quality (e.g., 3 bits for one of five possible bases and 5 bits for one of 32 quality values), FASTR uses a partitioned range encoding scheme over the 8-bit integer space. The full range of values from 0 to 255 is divided into five non-overlapping intervals, one for each nucleotide base: N, A, C, G, and T (or U in RNA). Within each base-specific interval, the offset from the lower bound represents the quality score. Formally, let $b \in \{N,A,C,G,T\}$ denote the nucleotide base and let q denote the scaled quality value. Each base is assigned a contiguous subrange $[L_b,U_b]$ within the 8-bit space. The encoded value v is computed as $v = L_b+q$, where $q \in [0,62]$. The mapping is deterministic and reversible. During decoding, the base identity is determined by the interval containing $v$, and the quality value is recovered as $q=v-L_b$. In the default configuration, the ranges are assigned as follows: N occupies values $[0,2]$, A occupies $[3,65]$, G occupies $[66,128]$, C occupies $[129,191]$, and T occupies $[192,254]$. The remaining value, $255$, is reserved for FASTR’s safe mode, which we explain in Section \ref{subsec:record-deliminating}. These assignments ensure that each base has sufficient capacity to represent the standard range of Phred quality scores. The N base is given a much smaller range (only 3 values) because it represents an unknown or ambiguous nucleotide, for which quality scores are typically uninformative and therefore do not require the full dynamic range. It can be sufficient to represent the base quality of the base N as low, medium, and high quality. 
In most of today's sequencing technologies, a single ASCII character (e.g., ! in PacBio) is always assigned to the base N.

\begin{figure}[!htp]
    \centering
    \resizebox{\columnwidth}{!}{%
        \begin{tabular}{cccccc}
        \rowcolor{violet}\color{white}\textbf{N} & \color{white}\textbf{A} & \color{white}\textbf{C} & \color{white}\textbf{G} & \color{white}\textbf{T or U} & \color{white}\textbf{} \\ \hline
        \multicolumn{1}{|c|}{\cellcolor{orange} 0 \dots 2} &\multicolumn{1}{c|}{\cellcolor{blue}\color{white} 3 \dots 65} &\multicolumn{1}{c|}{\cellcolor{green} 66 \dots 128} &\multicolumn{1}{c|}{\cellcolor{red}\color{white} 129 \dots 191} &\multicolumn{1}{c|}{\cellcolor{yellow} 192 \dots 254} &\multicolumn{1}{c|}{\cellcolor{black}\color{white} 255} \\ \hline
        \end{tabular}
    }
    \vspace{0.1em}
    \caption{FASTR’s implicit partitioned range encoding scheme within the 8-bit integer space. In a single byte, we have 255 (0–254) possible scalar values, each of which represents a unique pair of base and its quality score. FASTR reserves the value 255 as an end-of-header character for its safe operation.}
    \label{fig:fastr-encoding}
\end{figure}

\subsection{Quality Score Processing and Scaling}
\label{subsec:quality-scaling}

FASTR supports flexible and extensible processing of base quality scores to accommodate diverse sequencing technologies, Phred standards, and downstream analysis requirements. Since quality scores in FASTQ are stored as ASCII characters with varying offsets and alphabets, FASTR normalizes all quality values into a unified numeric representation prior to encoding. During preprocessing, FASTR converts each ASCII quality character to its corresponding numeric Phred value using the user-specified or inferred offset. The system supports widely used Phred alphabets, including Phred42, Phred63, Phred68Solexa, Phred94, and Illumina RTA3/RTA4 bins. This conversion ensures that quality values are represented as integers in a consistent domain regardless of the original encoding scheme.

After numeric conversion, FASTR optionally applies a user-defined scaling function, $Q$, to transform the original quality values into a compact range suitable for partitioned encoding. By default, quality values are linearly mapped to the range $[0,62]$. Users may specify alternative scaling strategies, including logarithmic scaling, custom logarithmic functions with tunable parameters, or fully custom mathematical expressions. These functions are evaluated per quality value and allow users to control the trade-off between resolution and compactness. The scaled value is then added to the base’s lower bound to produce the final encoded byte. If the scaling function produces a value outside this domain, it is clamped to the nearest valid boundary to prevent overflow beyond the assigned range. FASTR also supports minimum quality thresholding, allowing users to discard or cap values below a specified threshold. All quality processing parameters, including the Phred alphabet, offset, scaling method, and custom formula parameters, are stored in the FASTR file header. This ensures that quality values can be accurately and deterministically reconstructed during FASTR-to-FASTQ conversion.

\subsection{Record Delimiting and Lossless Stream Decoding}
\label{subsec:record-deliminating}

After encoding bases and quality scores into numerical vectors, FASTR must ensure that individual reads can be unambiguously separated in the binary stream. Unlike text formats, newline characters or line delimiters in binary streams can be easily confused with numeric values falling in the base-quality encoding range. We solve this by introducing a reserved Sentinel Value that is mathematically unreachable by any valid base-quality encoding. In the partitioned range-encoding scheme, the maximum possible encoded value is {\tt 254}, corresponding to the T (or U in RNA) base with the highest scaled quality score. The value {\tt 255 (0xFF)}, therefore, remains unused in the encoding space. FASTR designates this value as a dedicated record delimiter. By inserting {\tt0xFF} between consecutive encoded reads, FASTR guarantees that any occurrence of this value in the binary stream marks a read boundary. This enables decoders and parallel workers to segment the stream without inspecting the encoded data's internal structure or storing the length of each read sequence in advance. The delimiter introduces a fixed overhead of an additional byte per read record. For example, a dataset containing 50 million reads incurs approximately 50 MB of delimiter overhead, which is negligible relative to typical sequencing file sizes (tens of GBs). During FASTR-to-FASTQ conversion, the decoder scans for sentinel values to reconstruct individual reads and applies base and quality decoding. This design enables efficient chunk-based processing and supports multithreaded processing by allowing threads to independently identify read boundaries.

\section{Results}
\label{sec:results}

We implement the complete FASTR software ecosystem in Python 3.
We evaluate the four modes of FASTR against three different groups of tools and file formats that aim to improve or replace FASTQ.
The first group includes 4 alternatives to FASTQ format: FASTR, SAM, BAM (compressed SAM), and CRAM.
The second group includes 2 state-of-the-art specialized FASTQ compression tools, ReNANO~\citep{dufort2021renano} and SPRING~\citep{chandak2019spring}.
The third group includes 2 general-purpose compression tools, pigz~\citep{adler2023pigz} (multithreaded implementation of Gzip) and XZ~\citep{collin2025xz}.
For each tool, we evaluate output file size, execution time, and peak memory for both compression (or encoding) and decompression (or decoding). 
We run all experiments on the AMD HPC cluster, which features a 2.5 GHz AMD EPYC 7V13 64-core CPU and 64 GB of RAM. 
We use 16 CPU threads for all tools.
We generate SAM using minimap2~\citep{li2018minimap2}, by mapping FASTQ to the primary assembly GRCh38.p14 (patch release 14, accession: GCF\_000001405.40).
We generate BAM and CRAM from the generated SAM using Samtools~\citep{li2009sequence}.
We run ReNANO without providing a reference genome.
We enable the \emph{-l} parameter of SPRING for processing long reads (i.e., HiFi and ONT).
We run pigz~\citep{adler2023pigz} in three processing modes: \textit{fast}, \textit{best}, and \textit{Zopfli}, with compression levels of 1, 9, and 11, respectively.
We run XZ in two processing modes: \textit{fast} and \textit{best}, with compression levels of 0 and 9, respectively. All used command-lines are reproducible and available on GitHub.


\begin{table*}[htbp]
    \centering
    \small
    \setlength{\tabcolsep}{3pt} 
    \caption{Benchmark comparison of FASTR, SAM, BAM, and CRAM generated from Illumina, HiFi, and ONT FASTQ files.}
    \label{table:alternatives}
    \begin{tabular}{ll rrr r rrr}
        \toprule
        & & \multicolumn{4}{c}{\textbf{Compression/Encoding}} & \multicolumn{3}{c}{\textbf{Decompression/Decoding}} \\
        \cmidrule(lr){3-6} \cmidrule(lr){7-9}
        & & \textbf{Output Size} & \textbf{Comp.} & \textbf{Time} & \textbf{Peak Mem} & \textbf{Output Size} & \textbf{Time} & \textbf{Peak Mem} \\
        \textbf{} & \textbf{Method} & \textbf{(MB)} & \textbf{Ratio} & \textbf{(Sec)} & \textbf{(MB)} & \textbf{(MB)} & \textbf{(Sec)} & \textbf{(MB)} \\
        \midrule
        \multirow{7}{*}{Illumina} 
        & FASTR \textit{mode 0} & 12137.1 & 1.42 & \textbf{846.0} & \textbf{153.2} & \textbf{14170.4} & 390.68 & 2598.2 \\
        & FASTR \textit{mode 1} & 8689.3 & 1.99 & \textbf{991.2} & \textbf{155.9} & \textbf{14170.4} & 609.37 & 446.7 \\
        & FASTR \textit{mode 2} & 6656.0 & 2.60 & \textbf{979.5} & \textbf{154.8} & \textbf{14170.4} & 758.27 & 539.0 \\
        & FASTR \textit{mode 3} & 5481.1 & 3.15 & \textbf{895.6} & \textbf{173.2} & \textbf{14170.4} & 6180.52 & 862.2 \\
        & SAM & 18524.4 & 0.93 & 270466.0 & 15087.9 & 11923.5 & 106.68 & 29.1 \\
        & BAM & 2701.6 & 6.40 & 270774.8 & 15087.9 & 11923.5 & \textbf{100.68} & \textbf{11.2} \\
        & CRAM & \textbf{1361.8} & \textbf{12.70} & 271070.0 & 15087.9 & 11923.5 & 146.27 & 3633.8 \\
        \midrule
        \multirow{7}{*}{\makecell{PacBio \\ HiFi}} 
        & FASTR \textit{mode 0} & 5965.7 & 1.00 & \textbf{64.0} & \textbf{190.4} & \textbf{5972.1} & 53.52 & 415.8 \\
        & FASTR \textit{mode 1} & 2990.6 & 2.00 & \textbf{74.4} & \textbf{175.6} & \textbf{5972.1} & 58.36 & 463.6 \\
        & FASTR \textit{mode 2} & 2984.2 & 2.00 & \textbf{66.8} & \textbf{162.3} & \textbf{5972.1} & 55.02 & 1171.9 \\
        & FASTR \textit{mode 3} & 2981.5 & 2.01 & \textbf{68.0} & \textbf{176.0} & \textbf{5972.1} & 54.19 & 1120.5 \\
        & SAM & 6119.6 & 0.98 & 2349.4 & 22633.4 & 5965.5 & \textbf{24.65} & \textbf{64.7} \\
        & BAM & 1468.4 & 4.07 & 2518.7 & 22633.4 & 5965.5 & 30.44 & 12.3 \\
        & CRAM & \textbf{558.2} & \textbf{10.72} & 2602.7 & 22633.4 & 5965.5 & 29.66 & 4160.1 \\
        \midrule
        \multirow{7}{*}{ONT} 
        & FASTR \textit{mode 0} & 7996.1 & 1.00 & \textbf{88.3} & \textbf{198.6} & \textbf{8000.5} & 73.27 & 338.5 \\
        & FASTR \textit{mode 1} & 4006.4 & 2.00 & \textbf{107.0} & \textbf{166.1} & \textbf{8000.5} & 82.5 & 830.5 \\
        & FASTR \textit{mode 2} & 4002.0 & 2.00 & \textbf{91.4} & \textbf{167.2} & \textbf{8000.5} & 73.72 & 1472.9 \\
        & FASTR \textit{mode 3} & 3994.1 & 2.01 & \textbf{113.6} & \textbf{174.6} & \textbf{8000.5} & 80.4 & 556.7 \\
        & SAM & 8632.9 & 0.93 & 18961.0 & 19737.8 & 7991.6 & \textbf{34.54} & 346.4 \\
        & BAM & 3563.5 & 2.25 & 19124.1 & 19737.8 & 7991.6 & 48.01 & \textbf{15.4} \\
        & CRAM & \textbf{2074.1} & \textbf{3.86} & 19242.5 & 19737.8 & 7991.6 & 47.74 & 4441.9 \\
        \bottomrule
    \end{tabular}
\end{table*}

\subsection{Evaluated Datasets}
Our experimental evaluation uses 3 different real FASTQ read sets downloaded from SRA.
The first FASTQ file is generated by Illumina's NextSeq 2000, with accession number ERR15909551, a total file size of 17,290,229,774 bytes, and a read length of 122 bases.
The second FASTQ file is generated by PacBio's Revio (HiFi data), with accession number ERR13491966, a total file size of 5,980,918,708 bytes, and an average read length of 22,829 bases (the longest is 59,620 bases).
The third FASTQ file is generated by ONT's PromethION, with accession number SRR33464820, a total file size of 8,012,401,592 bytes, and an average read length of 22,362 bases (the longest is 664,420 bases).

\subsection{FASTR is the Best Successor to FASTQ}
Comparison of FASTR with existing FASTQ replacement formats, SAM, BAM, and CRAM as presented in Table \ref{table:alternatives}, provides four observations.
1) FASTR is on average 293$\times$, 38$\times$, 194$\times$ faster to generate from Illumina, HiFi, and ONT FASTQs, respectively, compared to generating alignment-based SAM, BAM, and CRAM.
2) FASTR is always 2-3.15$\times$ smaller in size compared to FASTQ, depending on the processing mode. 
Although most of the size reduction is due to our implicit partitioned range encoding approach, optimizing read headers alone yields a size reduction of up to 1.42$\times$.
3) CRAM is only 4$\times$, 5.3$\times$, and 1.9$\times$ smaller than FASTR using Illumina, HiFi, and ONT reads, respectively. But CRAM is still largely inefficient due to three main reasons: (1) CRAM requires an extremely long execution time compared to FASTR, (2) CRAM requires a reference genome, which can be unavailable, as in metagenomic FASTQs, novel species, and unknown sample donors, and (3) We observe that CRAM (also SAM and BAM) always encodes read headers in a lossy approach, as they retain only the substring preceding the first space character in the header.
4) Conversion from FASTR, SAM, BAM, and CRAM back to FASTQ is, in most cases, much faster than converting from FASTQ to these formats.

We conclude that FASTR stands out as the most practical successor to FASTQ by combining extremely fast generation, substantial size reduction, reference-free operation, and efficient conversion back to FASTQ.

\begin{table*}[htbp]
    \centering
    \setlength{\tabcolsep}{8pt}
    \caption{Compression and Decompression performance using FASTQ-specialized (ReNANO and SPRING) and general-purpose (pigz and XZ) compression tools, applied on Illumina, PacBio HiFi, and ONT data in FASTR and FASTQ formats.}
    \label{table:compression}
    \begin{tabular}{ll rrrr rrrr}
        \toprule
        & & \multicolumn{4}{c}{\textbf{Compression}} & \multicolumn{3}{c}{\textbf{Decompression}} \\
        \cmidrule(lr){3-6} \cmidrule(lr){7-9}
        & & \textbf{Size} & \textbf{Comp.} & \textbf{Time} & \textbf{Mem} & \textbf{Size} & \textbf{Time} & \textbf{Mem} \\
        \textbf{} & \textbf{Method} & \textbf{(MB)} & \textbf{Ratio} & \textbf{(sec)} & \textbf{(MB)} & \textbf{(MB)} & \textbf{(sec)} & \textbf{(MB)} \\
        \midrule
        \parbox[t]{2mm}{\multirow{12}{*}{\rotatebox[origin=c]{90}{Illumina}}} 
        & ReNANO & 1096.67 & 15.77 & 263.68 & 499.85 & 17290.23 & 439.76 & 496.53 \\
        & SPRING & \textbf{884.28} & \textbf{19.55} & 15887.30 & 6240.33 & 14170.40 & 1215.37 & 5386.59 \\
        & pigz(best)+FASTQ & 1949.58 & 8.87 & 3805.80 & \textbf{20.73} & 17290.23 & 65.25 & \textbf{4.11} \\
        & pigz(best)+FASTR & 1553.42 & 11.13 & 2588.18 & \textbf{18.43} & \textbf{6656.01} & \textbf{35.73} & \textbf{4.24} \\
        & pigz(fast)+FASTQ & 2704.56 & 6.39 & 152.89 & \textbf{19.04} & 17290.23 & 85.84 & \textbf{4.11} \\
        & pigz(fast)+FASTR & 2085.32 & 8.29 & \textbf{91.36} & \textbf{19.94} & \textbf{6656.01} & \textbf{36.73} & \textbf{4.12} \\
        & pigz(Zopfli)+FASTQ & 1831.36 & 9.44 & 88610.30 & 129.00 & 17290.23 & 69.22 & \textbf{4.12} \\
        & pigz(Zopfli)+FASTR & 1456.03 & 11.87 & 35818.00 & 153.28 & \textbf{6656.01} & \textbf{34.76} & \textbf{4.24} \\
        & XZ(best)+FASTQ & 1400.95 & 12.34 & 13402.80 & 14627.10 & 17290.23 & 123.13 & 6987.89 \\
        & XZ(best)+FASTR & 1102.91 & 15.68 & 10486.10 & 15049.32 & \textbf{6656.01} & 83.38 & 7264.98 \\
        & XZ(fast)+FASTQ & 1971.60 & 8.77 & 665.31 & 214.80 & 17290.23 & 125.97 & 110.66 \\
        & XZ(fast)+FASTR & 1598.98 & 10.81 & 502.23 & 233.05 & \textbf{6656.01} & 77.55 & 114.84 \\
        \midrule
        \parbox[t]{2mm}{\multirow{12}{*}{\rotatebox[origin=c]{90}{PacBio HiFi}}} 
        & ReNANO & 1201.81 & 4.98 & 145.43 & 598.46 & 5980.92 & 247.40 & 571.26 \\
        & SPRING & \textbf{1155.05} & \textbf{5.18} & 709.45 & 11154.61 & 5972.07 & 851.78 & 8869.13 \\
        & pigz(best)+FASTQ & 1538.18 & 3.89 & 3913.63 & \textbf{20.09} & 5980.92 & 33.76 & \textbf{4.11} \\
        & pigz(best)+FASTR & 1323.25 & 4.52 & 1265.96 & \textbf{21.25} & \textbf{2984.19} & \textbf{20.86} & \textbf{4.24} \\
        & pigz(fast)+FASTQ & 1823.16 & 3.28 & 99.39 & \textbf{17.64} & 5980.92 & 35.02 & \textbf{4.12} \\
        & pigz(fast)+FASTR & 1446.13 & 4.14 & \textbf{63.33} & \textbf{18.64} & \textbf{2984.19} & \textbf{21.82} & \textbf{4.24} \\
        & pigz(Zopfli)+FASTQ & 1340.79 & 4.46 & 59739.10 & 114.21 & 5980.92 & 38.58 & \textbf{4.12} \\
        & pigz(Zopfli)+FASTR & 1234.09 & 4.85 & 16395.40 & 184.42 & \textbf{2984.19} & \textbf{19.70} & \textbf{4.24} \\
        & XZ(best)+FASTQ & 1242.55 & 4.81 & 8343.49 & 15303.29 & 5980.92 & 96.65 & 6380.68 \\
        & XZ(best)+FASTR & \textbf{1160.40} & \textbf{5.15} & 7528.29 & 14158.61 & \textbf{2984.19} & 62.75 & 4516.71 \\
        & XZ(fast)+FASTQ & 1518.55 & 3.94 & 392.08 & 226.22 & 5980.92 & 85.05 & 116.96 \\
        & XZ(fast)+FASTR & 1333.63 & 4.48 & 379.27 & 244.60 & \textbf{2984.19} & 55.42 & 121.33 \\
        \midrule
        \parbox[t]{2mm}{\multirow{12}{*}{\rotatebox[origin=c]{90}{ONT}}} 
        & ReNANO & 2994.70 & 2.68 & 217.77 & 647.87 & 8012.40 & 409.82 & 597.64 \\
        & SPRING & \textbf{2688.84} & \textbf{2.98} & 1137.97 & 13597.32 & 8000.46 & 950.25 & 9823.48 \\
        & pigz(best)+FASTQ & 3517.65 & 2.28 & 2891.25 & \textbf{19.19} & 8012.40 & 61.98 & \textbf{4.11} \\
        & pigz(best)+FASTR & 2808.54 & 2.85 & 605.97 & \textbf{19.84} & \textbf{4002.01} & \textbf{35.48} & \textbf{4.12} \\
        & pigz(fast)+FASTQ & 3809.00 & 2.10 & 145.84 & \textbf{20.84} & 8012.40 & 59.62 & \textbf{4.12} \\
        & pigz(fast)+FASTR & 2851.69 & 2.81 & \textbf{131.35} & \textbf{21.62} & \textbf{4002.01} & \textbf{36.82} & \textbf{4.11} \\
        & pigz(Zopfli)+FASTQ & 3210.65 & 2.50 & 59813.70 & 189.50 & 8012.40 & 59.44 & \textbf{4.12} \\
        & pigz(Zopfli)+FASTR & \textbf{2702.29} & \textbf{2.97} & 13970.50 & 253.99 & \textbf{4002.01} & \textbf{34.23} & \textbf{4.12} \\
        & XZ(best)+FASTQ & 3002.43 & 2.67 & 17940.00 & 16546.08 & 8012.40 & 167.21 & 5343.44 \\
        & XZ(best)+FASTR & 2466.49 & 3.25 & 6623.20 & 16596.38 & \textbf{4002.01} & 161.51 & 5653.86 \\
        & XZ(fast)+FASTQ & 3380.92 & 2.37 & 970.37 & 259.06 & 8012.40 & 143.90 & 107.42 \\
        & XZ(fast)+FASTR & \textbf{2675.84} & \textbf{2.99} & 975.91 & 249.24 & \textbf{4002.01} & 98.32 & 129.59 \\
        \bottomrule
    \end{tabular}%
\end{table*}

\subsection{FASTR Boosts Compression Performance}
We evaluate the use of specialized and general-purpose compression tools with FASTR and FASTQ.
Based on Table \ref{table:compression}, we make four key observations.
1) FASTR consistently improves the execution time of the state-of-the-art general-purpose compression tools, pigz and XZ, by 2.47$\times$, 3.64$\times$, and 4.8$\times$ using Illumina, HiFi, and ONT reads, respectively, compared to using them with FASTQ files.
2) Using FASTR files, pigz and XZ consistently provide higher compression ratios (1.1-1.3$\times$) than that with using FASTQ files. 
For example, pigz(\textit{best}) and XZ(\textit{best}) with Illumina data provide a compression ratio of 11.13$\times$ and 15.68$\times$, respectively, when using FASTR compared to only 8.87$\times$ and 12.34$\times$, respectively, when using FASTQ.
3) FASTR consistently improves the decompression time of pigz and XZ by 2.34$\times$, 1.96$\times$, 1.75$\times$ when using Illumina, HiFi, and ONT reads, respectively, compared to that with using FASTQ files.
This, for example, allows pigs with FASTR to provide 35$\times$, 43.2$\times$, and 27.7$\times$ (12.65$\times$, 12.6$\times$, 12$\times$) less decompression time compared to SPRING (and ReNANO).
4) We observe that SPRING provides about 1.25$\times$ higher compression ratio than using XZ(\textit{best}) with FASTR and Illumina data. For HiFi and ONT data, SPRING and ReNANO provide almost the same compression ratios as with pigz and XZ with FASTR.
However, both ReNANO and SPRING suffer from an extremely higher compression time (1-2 orders of magnitude), decompression time (up to an order of magnitude), and peak memory (2-3 orders of magnitude), compared to pigz(\textit{fast}) with FASTR.
We observe that SPRING does \emph{not} reconstruct back the second header (following the '+' sign) of the FASTQ, causing the decompressed FASTQ to be smaller in size.
FASTR decoding does the same by default, but users can use the \textit{\texttt{--second\_head}} parameter to losslessly reconstruct the second header.

We conclude that FASTR substantially accelerates compression and decompression while improving compression ratios, enabling general-purpose compression tools to rival specialized compressors with far lower time and memory overhead.

\subsection{FASTR is Directly Usable Without Decompression}
FASTR represents sequencing reads in compact numeric values, yet directly interpretable form, allowing FASTQ-compatible tools to process FASTR files without prior decompression.
We demonstrate this benefit by enabling the most widely used read mapping tool, minimap2~\citep{li2018minimap2}, to use FASTR file format.
We only add about 20 lines of code to minimap2 (bseq.c and kseq.h) to allow minimap2 to detect our reserved sentinel value to correctly parse each numeric vector and use lookup table operation to quickly map each numeric value to its base. 
Table \ref{table:minimap2} demonstrates that FASTR format does \emph{not} introduce any computational overhead to minimap2~\citep{li2018minimap2}.
FASTR even reduces the associated I/O costs, resulting in a slight reduction in total execution time.
This can be extremely important for applications, including minimap2, that show slow I/O performance (e.g., piping to gzip, or reading from a slow pipe).
As FASTR optimizes the read headers (e.g., storing the accession number only once in the file header), the first column of the SAM file has to maintain only the read identifier.
This results in an additional reduction in the SAM file size.

We conclude that FASTR enables drop-in compatibility with existing tools like minimap2, reducing I/O and output sizes without incurring any computational overhead.

\begin{table}[htbp]
    \centering
    \setlength{\tabcolsep}{3pt} 
    \caption{Minimap2 mapping performance after using FASTR versus FASTQ across Illumina, PacBio HiFi, and ONT data.}
    \resizebox{\columnwidth}{!}{
    \label{table:minimap2}
    \begin{tabular}{llrrr}
        \toprule
        \textbf{} & \textbf{Method} & \textbf{\begin{tabular}[c]{@{}c@{}} SAM Size \\(MB)\end{tabular}} & \textbf{\begin{tabular}[c]{@{}c@{}}Mapping\\ Time (Sec)\end{tabular}} & \textbf{\begin{tabular}[c]{@{}c@{}} Peak Memory\\ (MB)\end{tabular}} \\
        \midrule
        \multirow{2}{*}{Illumina} & +FASTR & \textbf{12314.20} & 270520.0 & 15028.7 \\
                                 & +FASTQ & 12933.73 & \textbf{269912.0} & \textbf{14958.2} \\
        \midrule
        \multirow{2}{*}{PacBio HiFi}     & +FASTR & \textbf{3101.41} & \textbf{2251.0} & 22340.4 \\
                                 & +FASTQ & 3103.43 & 2290.2 & \textbf{22191.8} \\
        \midrule
        \multirow{2}{*}{ONT}      & +FASTR & \textbf{4522.33} & \textbf{18186.3} & \textbf{19018.1} \\
                                 & +FASTQ & 4526.25 & 18789.0 & 19410.1 \\
        \bottomrule
    \end{tabular}%
    }
\end{table}

\section{Conclusion}
FASTQ has served the sequencing community well by providing a simple and human-readable representation of reads. 
However, decades of advances in sequencing technologies and rapidly growing data volumes have rendered FASTQ increasingly inefficient. 
In this work, we reimagine what an alternative could be and introduce FASTR, a practical and efficient successor that eliminates interpretation ambiguity, reduces storage footprint, improves compression performance, and preserves essential experimental metadata. 
We provide a complete software ecosystem supporting FASTQ–FASTR conversion and multiple processing modes to accommodate diverse use cases. 
Our results demonstrate that FASTR can be adopted by downstream tools with minimal interface-level changes and no modification to core algorithms. 
By combining efficiency, compatibility, and extensibility, FASTR lays the foundation for a new generation of sequencing tools and data workflows.
Examples include quality control tools, specialized compression tools, visualization tools, parsers for different programming languages, and many more.
As sequencing technologies continue to improve sequencing accuracy, the effective range of base quality values is expected to shrink. 
This creates new opportunities to encode additional information within a single byte using our partitioned range encoding scheme, further enhancing the FASTR format and enabling support for a broader range of genomic data types.
There is also more room for improving FASTR software by re-implementing it in the C programming language.
We hope that these efforts and the challenges we discuss provide a foundation for future work to catalyze existing genomic formats and enable new applications.

\section{Competing interests}
No competing interest is declared.

\section{Author contributions statement}
M.A. led the project, 
M.A. conceived of the presented idea, 
A.T. developed FASTR in Python, 
M.A. integrated FASTR with minimap2,
S.S., Z.B., and M.A. performed the statistical analysis and produced the figures and tables,
A.E.A., M.N.U., A.T., and M.A. created scripts for running and evaluating software tools,
A.P. developed the website and the online visualization tool,
A.T., S.S., Z.B., A.Z., S.M., C.A., and M.A. wrote, reviewed, and edited the manuscript,
All authors read and approved the final manuscript.

\section{Acknowledgments}
M.A. acknowledges a startup fund from Georgia State University (GSU), provided by the Department of Computer Science and the College of Arts and Sciences. 
M.A. also acknowledges a Faculty Internal Grant (FIG) from Georgia State University (GSU), project "MetaGo: AI-Driven Reproducible Context and Sample -Aware Metagenomics Analyses", and an industrial grant from Advanced Micro Devices (AMD), Inc.
S.M. and A.Z., were supported by a grant of the Ministry of Research, Innovation and Digitization, under the Romania’s National Recovery and Resilience Plan – Funded by EU – NextGenerationEU program, project "Metagenomics and Bioinformatics tools for Wastewater-based Genomic Surveillance of viral Pathogens for early prediction of public health risks – (MetBio-WGSP)" number 760286/27.03.2024, code 167/31.07.2023, within Pillar III, Component C9, Investment 8.
C.A. and Z.B., were supported by the European Union’s Horizon Programme for Research and Innovation under grant agreement No. 101047160, project BioPIM (awarded to C.A.).

\bibliographystyle{abbrvnat}
\bibliography{bibtex}

\begin{thebibliography}{19}
\providecommand{\natexlab}[1]{#1}
\providecommand{\url}[1]{\texttt{#1}}
\expandafter\ifx\csname urlstyle\endcsname\relax
  \providecommand{\doi}[1]{doi: #1}\else
  \providecommand{\doi}{doi: \begingroup \urlstyle{rm}\Url}\fi

\bibitem[Adler(2023)]{adler2023pigz}
M.~Adler.
\newblock pigz: A parallel implementation of gzip for modern multi-processor,
  multi-core machines.
\newblock \url{https://zlib.net/pigz/}, 2023.
\newblock Version 2.8.

\bibitem[Alser et~al.(2022)Alser, Lindegger, Firtina, Almadhoun, Mao, Singh,
  Gomez-Luna, and Mutlu]{alser2022molecules}
M.~Alser, J.~Lindegger, C.~Firtina, N.~Almadhoun, H.~Mao, G.~Singh,
  J.~Gomez-Luna, and O.~Mutlu.
\newblock From molecules to genomic variations: Accelerating genome analysis
  via intelligent algorithms and architectures.
\newblock \emph{Computational and Structural Biotechnology Journal},
  20:\penalty0 4579--4599, 2022.

\bibitem[Alser et~al.(2025)Alser, Eudine, and Mutlu]{alser2025taming}
M.~Alser, J.~Eudine, and O.~Mutlu.
\newblock Taming large-scale genomic analyses via sparsified genomics.
\newblock \emph{Nature Communications}, 16\penalty0 (1):\penalty0 876, 2025.

\bibitem[Bray et~al.(2016)Bray, Pimentel, Melsted, and Pachter]{bray2016near}
N.~L. Bray, H.~Pimentel, P.~Melsted, and L.~Pachter.
\newblock {Near-Optimal Probabilistic RNA-seq Quantification}.
\newblock \emph{Nature Biotechnology}, 34\penalty0 (5):\penalty0 525--527,
  2016.

\bibitem[Brown and Irber(2016)]{brown2016sourmash}
C.~T. Brown and L.~Irber.
\newblock {Sourmash: a Library for MinHash Sketching of DNA}.
\newblock \emph{Journal of Open Source Software}, 1\penalty0 (5):\penalty0 27,
  2016.

\bibitem[Chandak et~al.(2019)Chandak, Tatwawadi, Ochoa, Hernaez, and
  Weissman]{chandak2019spring}
S.~Chandak, K.~Tatwawadi, I.~Ochoa, M.~Hernaez, and T.~Weissman.
\newblock {SPRING: a Next-Generation Compressor for FASTQ Data}.
\newblock \emph{Bioinformatics}, 35\penalty0 (15):\penalty0 2674--2676, 2019.

\bibitem[Cock et~al.(2010)Cock, Fields, Goto, Heuer, and Rice]{cock2010sanger}
P.~J. Cock, C.~J. Fields, N.~Goto, M.~L. Heuer, and P.~M. Rice.
\newblock {The Sanger FASTQ File Format for Sequences with Quality Scores, and
  the Solexa/Illumina FASTQ Variants}.
\newblock \emph{{Nucleic Acids Research}}, 38\penalty0 (6):\penalty0
  1767--1771, 2010.

\bibitem[Collin et~al.(2025)]{collin2025xz}
L.~Collin et~al.
\newblock {XZ Utils: A Free General-Purpose Data Compression Software}.
\newblock \url{https://tukaani.org/xz/}, 2025.
\newblock Version 5.8.2.

\bibitem[Deorowicz(2020)]{deorowicz2020fqsqueezer}
S.~Deorowicz.
\newblock {FQSqueezer: k-mer-based Compression of Sequencing Data}.
\newblock \emph{Scientific Reports}, 10\penalty0 (1):\penalty0 578, 2020.

\bibitem[Dufort~y {\'A}lvarez et~al.(2021)Dufort~y {\'A}lvarez, Seroussi,
  Smircich, Sotelo-Silveira, Ochoa, and Mart{\'\i}n]{dufort2021renano}
G.~Dufort~y {\'A}lvarez, G.~Seroussi, P.~Smircich, J.~Sotelo-Silveira,
  I.~Ochoa, and {\'A}.~Mart{\'\i}n.
\newblock {RENANO: a REference-based compressor for NANOpore FASTQ files}.
\newblock \emph{Bioinformatics}, 37\penalty0 (24):\penalty0 4862--4864, 2021.

\bibitem[Fritz et~al.(2011)Fritz, Leinonen, Cochrane, and
  Birney]{fritz2011efficient}
M.~H.-Y. Fritz, R.~Leinonen, G.~Cochrane, and E.~Birney.
\newblock {Efficient Storage of High Throughput DNA Sequencing Data Using
  Reference-based Compression}.
\newblock \emph{Genome Research}, 21\penalty0 (5):\penalty0 734--740, 2011.

\bibitem[Hach et~al.(2012)Hach, Numanagi{\'c}, Alkan, and
  Sahinalp]{hach2012scalce}
F.~Hach, I.~Numanagi{\'c}, C.~Alkan, and S.~C. Sahinalp.
\newblock {SCALCE: Boosting Sequence Compression Algorithms Using Locally
  Consistent Encoding}.
\newblock \emph{Bioinformatics}, 28\penalty0 (23):\penalty0 3051--3057, 2012.

\bibitem[Katz et~al.(2022)Katz, Shutov, Lapoint, Kimelman, Brister, and
  O’Sullivan]{katz2022sequence}
K.~Katz, O.~Shutov, R.~Lapoint, M.~Kimelman, J.~R. Brister, and
  C.~O’Sullivan.
\newblock {The Sequence Read Archive: a Decade More of Explosive Growth}.
\newblock \emph{{Nucleic Acids Research}}, 50\penalty0 (D1):\penalty0
  D387--D390, 2022.

\bibitem[Leinonen et~al.(2011)Leinonen, Sugawara, Shumway, and
  Collaboration]{leinonen2011sequence}
R.~Leinonen, H.~Sugawara, M.~Shumway, and I.~N. S.~D. Collaboration.
\newblock The sequence read archive.
\newblock \emph{Nucleic Acids Research}, 39\penalty0 (suppl\_1):\penalty0
  D19--D21, 2011.

\bibitem[Li(2018)]{li2018minimap2}
H.~Li.
\newblock Minimap2: pairwise alignment for nucleotide sequences.
\newblock \emph{Bioinformatics}, 34\penalty0 (18):\penalty0 3094--3100, 2018.

\bibitem[Li(2020)]{Li2020FastLanguages}
H.~Li.
\newblock Fast high-level programming languages.
\newblock
  \url{https://lh3.github.io/2020/05/17/fast-high-level-programming-languages},
  May 2020.
\newblock Accessed: 2024-05-20.

\bibitem[Li et~al.(2009)Li, Handsaker, Wysoker, Fennell, Ruan, Homer, Marth,
  Abecasis, and Durbin]{li2009sequence}
H.~Li, B.~Handsaker, A.~Wysoker, T.~Fennell, J.~Ruan, N.~Homer, G.~Marth,
  G.~Abecasis, and R.~Durbin.
\newblock {The Sequence Alignment/Map Format and SAMtools}.
\newblock \emph{Bioinformatics}, 25\penalty0 (16):\penalty0 2078--2079, 2009.

\bibitem[Liu et~al.(2025)Liu, Rodriguez, Munteanu, Ronkowski, Sharma, Alser,
  Andreace, Blekhman, B{\l}aszczyk, Chikhi, et~al.]{liu2025analysis}
S.~Liu, J.~S. Rodriguez, V.~Munteanu, C.~Ronkowski, N.~K. Sharma, M.~Alser,
  F.~Andreace, R.~Blekhman, D.~B{\l}aszczyk, R.~Chikhi, et~al.
\newblock Analysis of metagenomic data.
\newblock \emph{Nature Reviews Methods Primers}, 5\penalty0 (1):\penalty0 5,
  2025.

\bibitem[Meyer et~al.(2022)Meyer, Fritz, Deng, Koslicki, Lesker, Gurevich,
  Robertson, Alser, Antipov, Beghini, et~al.]{meyer2022critical}
F.~Meyer, A.~Fritz, Z.-L. Deng, D.~Koslicki, T.~R. Lesker, A.~Gurevich,
  G.~Robertson, M.~Alser, D.~Antipov, F.~Beghini, et~al.
\newblock Critical assessment of metagenome interpretation: the second round of
  challenges.
\newblock \emph{Nature methods}, 19\penalty0 (4):\penalty0 429--440, 2022.

\end{thebibliography}




\end{document}